\documentclass[a4paper,10pt,oneside,aps,pra,notitlepage,nofootinbib,superscriptaddress,tightenlines,longbibliography]{revtex4-1}

\usepackage[utf8]{inputenc}
\usepackage[colorlinks,citecolor=black,linktocpage,breaklinks,hypertexnames=false,pdfpagelabels,draft=false]{hyperref}

\usepackage[intlimits]{amsmath}
\usepackage{amssymb, amsfonts,amsthm}
\usepackage{enumerate}
\usepackage{quantum}
\usepackage[capitalise]{cleveref}
\usepackage[dvips]{graphicx}

\setcitestyle{authoryear}

\theoremstyle{plain}
\newtheorem{lemma}{Lemma}
\newtheorem{thm}[lemma]{Result}

\newtheorem{idea}[lemma]{Idea}

\newtheorem{problem}[lemma]{Problem}

\renewcommand{\epsilon}{\varepsilon}

\newcommand{\be}{\begin{equation}}
\newcommand{\ee}{\end{equation}}
\newcommand{\bea}{\begin{eqnarray}}
\newcommand{\eea}{\end{eqnarray}}

\usepackage{todonotes}

\begin{document}

%\listoftodos

%\clearpage

\title{Comment on ``On the uncomputability of the spectral gap''}

\author{T. S. Cubitt}
\affiliation{Department of Computer Science, University College London, Gower Street, London WC1E 6BT, UK}

\author{D. P\'erez-Garc\'ia}
\affiliation{Departamento de An\'alisis Matem\'atico and IMI, Universidad Complutense de Madrid, 28040 Madrid, Spain\\
and ICMAT, C/ Nicol\'as Cabrera, Campus de Cantoblanco, 28049 Madrid, Spain}

\author{M. M. Wolf}
\affiliation{Zentrum Mathematik, Technische Universit\"at M\"unchen, 85748 Garching, Germany}

\begin{abstract}
The aim of this short note is to clarify some of the claims made in the comparison made in [S. Lloyd, {\it On the uncomputability of the spectral gap}, arXiv:1602.05924] between our recent result [T.S. Cubitt, D. Perez-Garcia, M.M. Wolf, {\it Undecidability of the spectral gap}, Nature 528, 207-211 (2015), arXiv:1502.04573] and his 1994 paper [S. Lloyd, {\it Necessary and sufficient conditions for quantum computation}, J. Mod. Opt. 41(12), 2503-2520 (1994)].
\end{abstract}

\maketitle

%\tableofcontents

In his recent paper \citep{Lloyd16}, Lloyd claims that in his 1993-1994 papers \citep{Lloyd93,Lloyd94} he already proved undecidability of the spectral gap problem for quantum Hamiltonians. We disagree with this.
Before detailing briefly the contribution of his and our paper to justify this assertion, it is important to specify first what the spectral gap is, and what is the spectral gap problem.  The spectral gap we are interested in is the difference between the two smallest energy levels in a Hamiltonian. The general spectral gap problem can  then be stated as:
\begin{problem}[General spectral gap problem]\label{spectral-gap-general}
Given a quantum Hamiltonian, does it have a spectral gap or not?
\end{problem}
If one restricts the Hamiltonians in the problem to translationally invariant ones given by nearest neighbor interactions on a  2D lattice, one obtains the problem that we address in our work:
\begin{problem}[Spectral gap problem for quantum many-body systems]\label{spectral-gap-ours}
Consider particles with a fixed and finite Hilbert space dimension on a 2D square lattice. Given a nearest neighbor interaction between them, does the associated translationally invariant Hamiltonian have spectral gap or not when the number of particles grows to infinity?
\end{problem}
Our result, proven in \citep{Cubitt15a,Cubitt15b}, is:
\begin{thm}[Our result]
  Problem \ref{spectral-gap-ours} is undecidable.
\end{thm}
\noindent (This also immediately implies that Problem \ref{spectral-gap-general} is undecidable.)

Problem \ref{spectral-gap-ours} lies at the core of condensed matter physics since it is behind the definition and understanding of quantum phases and quantum phase transitions. This means our result has striking consequences: the existence of quantum spin systems on the lattice with a huge degree of non-robustness. Or the existence of a new physical behavior in quantum many-body systems that we named ``size-driven phase transitions'' \citep{Bausch15}; systems for which any numerical approach to understand their low energy behavior in the thermodynamic limit will fail since an unpredictable transition to totally different properties may hold at any (arbitrarily large) system size. Neither of these consequences could have been derived from undecidability of Problem \ref{spectral-gap-general} (nor from Lloyd's Result \ref{thm:Seth-H}, which we explain now).

The problem that was considered and shown undecidable by Lloyd in \citep{Lloyd93,Lloyd94}, though undoubtedly interesting, is a different one. It is argued there that the unitary evolution $U$ associated to the evolution of a computer (classical or quantum) capable of universal computation has invariant subspaces with discrete spectrum (roots of unity) and other invariant subspaces with continuous spectrum (the whole unit circle), corresponding respectively to computations that halt and do not halt. As explained in \citep{Lloyd16}, the very nice idea behind that result is that:
\begin{idea}
\label{Seth-idea}
Computations that halt only explore a finite region of the Hilbert space while computations that do not halt explore an infinite region of the Hilbert space.
\end{idea}

Then, since the halting problem is undecidable,
\begin{thm}[Lloyd's result]\label{thm:Seth}
Given a quantum state in the infinite dimensional space in which $U$ is defined, it is undecidable to know whether it has overlap with an invariant subspace having discrete spectrum or, on the contrary, it is supported on an invariant subspace in which the spectrum is the full unit circle.
\end{thm}
\noindent (See \citep{Lloyd93}: ``there is no effective procedure that allows one to discover whether an arbitrary program state $\ket{b}$ is to be decomposed in terms of the discrete or the continuous part of the spectrum.'';
\citep{Lloyd94}: ``the answer to the question, whether the quantum computational state associated with a program falls in the discrete or continuous part of the spectrum, is uncomputable.'')

If, as commented in \citep{Lloyd16}, one considers the Hamiltonian $H=U+U^\dagger$, the same result holds true for a Hamiltonian instead of a unitary:
\begin{thm}\label{thm:Seth-H}
Given a quantum state in the infinite-dimensional space in which $H$ is defined, it is undecidable to know whether it has overlap with an invariant subspace having discrete spectrum or, on the contrary, it is supported on an invariant subspace in which the spectrum is $[-2,2]$.
\end{thm}
\noindent Note that the spectrum of $H$ is always $[-2,2]$ and therefore its spectral gap, as defined above, is always~$0$ and not undecidable. Note also that the full spectral decomposition -- including invariant subspaces -- appears in the problem (as opposed to the spectral gap problem where just the two smallest eigenvalues are involved). Finally note that the problem is about the overlap of a given quantum state, which is the input to the problem, with the invariant subspaces of a fixed unitary (or Hamiltonian).

Let us conclude by saying that, contrary to what is claimed in \citep{Lloyd16}, our result is not based on Idea \ref{Seth-idea}. The ideas behind our proof are clearly explained in \citep{Cubitt15a}, and Idea \ref{Seth-idea} is not used in any way, direct or indirect, in our derivation.

%%===========================

\bibliography{lloyd-comment-biblio}

\end{document}